\documentclass[fp,twocolumn]{jpsj3}
\usepackage{txfonts}
\usepackage{bm}
\usepackage{multirow}

\title{Mpemba effect in a sheared granular gas with velocity-dependent restitution}

\author{
    Makoto R. Kikuchi$^1$\thanks{m-kikuchi@st.go.tuat.ac.jp},
    Yuria Kobayashi$^1$, and 
    Satoshi Takada$^1$\thanks{Corresponding author, takada@go.tuat.ac.jp}}
\inst{
    $^1$Department of Mechanical Systems Engineering,
    Tokyo University of Agriculture and Technology,\\
    2-24-16 Naka-cho, Koganei, Tokyo 184-8588, Japan} %\\

\abst{
We investigate the Mpemba effect in a dilute sheared granular gas with a velocity-dependent restitution coefficient. 
Using kinetic theory based on Grad's moment method, we analyze the relaxation dynamics following a sudden change in the shear rate. 
We show that, despite having a higher initial temperature, a system starting from an isotropic state can relax faster than a system prepared in a sheared steady state, demonstrating a clear Mpemba effect in the temperature evolution. 
We further demonstrate the emergence of a viscosity Mpemba effect, characterized by crossings in the relaxation curves of the shear viscosity. 
Remarkably, multiple crossings arise due to an additional intrinsic timescale introduced by the velocity dependence of the restitution coefficient, providing a minimal kinetic mechanism for multiple Mpemba effects in driven granular gases.
}

\begin{document}
\maketitle
\section{Introduction}
The Mpemba effect~\cite{Mpemba1969, Burridge16, Katz17, Lu17, Kumar20, Kumar22, Santos24, Santos26} refers to the counterintuitive relaxation phenomenon in which a system initially prepared at a higher temperature reaches equilibrium faster than an identical system prepared at a lower temperature. 
Originally observed in water freezing~\cite{Mpemba1969, Burridge16, Katz17}, it is now recognized as a generic feature~\cite{Lu17, Klich19, Busiello21} of nonequilibrium relaxation arising from the coexistence of multiple relaxation modes and their dependence on initial conditions.

Granular gases or suspensions provide a particularly suitable platform for studying the Mpemba effect because of their intrinsic dissipation and rich nonequilibrium dynamics. 
Mpemba-like effects have been reported in freely cooling, driven, and sheared systems~\cite{Lasanta17, Torrente19, Biswas20, Mompo21, Takada21, Patron23}, where relaxation is influenced not only by energy dissipation but also by additional internal degrees of freedom such as stress and velocity anisotropy.

The Mpemba effect has also been discussed in a broader context, including driven classical and quantum systems~\cite{Baity19, Carollo21, Strachan25, Ares25Rev, Chatterjee23, Chatterjee24}.
This suggests that it represents a universal aspect of nonequilibrium relaxation.

Despite these advances, the role of rheological effects and velocity-dependent dissipation in generating Mpemba behavior in granular gases remains largely unexplored~\cite{Takada21}. 
In particular, a velocity-dependent restitution coefficient introduces an intrinsic velocity scale, leading to additional relaxation times even in dilute systems~\cite{Scheffler02, Poschel03, Takada17, Takada22, Kikuchi26}. 
Here, we show that shear flow qualitatively modifies relaxation pathways and leads not only to a temperature Mpemba effect but also to a viscosity Mpemba effect~\cite{Takada21}. 
Furthermore, the additional intrinsic timescale gives rise to multiple crossings in relaxation curves, which are absent in simpler models.
The present study is also related to our previous work on discontinuous shear rheology in similar systems; however, here we focus on transient relaxation dynamics and Mpemba effects, which are fundamentally different from steady-state rheology~\cite{Kikuchi26}.

In this paper, we investigate the Mpemba effect in a dilute sheared granular gas with a velocity-dependent restitution coefficient. 
Using kinetic theory~\cite{Chapman, Brilliantov, Garzo} based on Grad's moment method~\cite{Grad49}, we analyze the relaxation dynamics following a sudden change in the shear rate and demonstrate both temperature and viscosity Mpemba effects.

The remainder of this paper is organized as follows.
In Sec.~\ref{sec:model}, we introduce the model and simulation setup.
In Sec.~\ref{sec:kinetic_theory}, we present the kinetic-theoretical formulation based on Grad's moment method.
In Sec.~\ref{sec:protocol}, we define the Mpemba protocols and specify the initial conditions.
The results are presented in Sec.~\ref{sec:results}, where we demonstrate the emergence of both temperature and viscosity Mpemba effects.
In Sec.~\ref{sec:discussion}, we discuss the assumptions, limitations, and physical implications of the present approach.
Finally, Sec.~\ref{sec:summary} summarizes the main findings of this study.

%%%%%%%%%%%%%%%%%%%%%%%%%%%%%%%%%
\section{Model and setup}\label{sec:model}
We study a dilute granular gas composed of identical, frictionless hard spheres of mass $m$ and diameter $d$.
The system is assumed to be sufficiently rarefied so that binary collisions dominate, and the solid volume fraction is fixed at $\varphi=0.01$.
The position and velocity of particle $i$ are denoted by $\bm{r}_i$ and $\bm{v}_i$, respectively.
For a binary collision between two particles with pre-collisional velocities $(\bm{v}_1, \bm{v}_2)$, the post-collisional velocities are given by
\begin{equation}
    \bm{v}_1^\prime = \bm{v}_1 - \frac{1+e}{2}\left(\bm{v}_{12}\cdot \hat{\bm{k}}\right)\hat{\bm{k}},\quad
    \bm{v}_2^\prime = \bm{v}_2 + \frac{1+e}{2}\left(\bm{v}_{12}\cdot \hat{\bm{k}}\right)\hat{\bm{k}},
    \label{eq:pre_post_v}
\end{equation}
where $\bm{v}_{12}\equiv \bm{v}_1-\bm{v}_2$, and $\hat{\bm{k}}\equiv (\bm{r}_1-\bm{r}_2)/|\bm{r}_1-\bm{r}_2|$ is the unit vector pointing along the line of centers at contact.
The normal restitution coefficient is assumed to depend on the normal component $v_n$ of the relative velocity according to
\begin{equation}
    e(v_n)=e_1 - (e_1-e_2)\Theta(v_n-v_\mathrm{c}),
    \label{eq:def_COR}
\end{equation}
where $v_\mathrm{c}$ represents a characteristic threshold velocity.
Velocity-dependent restitution coefficients have been widely studied in the context of viscoelastic and charged granular gases~\cite{Scheffler02, Poschel03, Takada17, Takada22, Kikuchi26}, where the restitution coefficient naturally depends on the impact velocity.
Although the stepwise form adopted in Eq.~\eqref{eq:def_COR} is a simplified representation, it captures the essential feature of introducing a characteristic velocity scale into the system.
Therefore, the present model can be regarded as a minimal framework to investigate the effect of velocity-dependent dissipation on nonequilibrium relaxation.

A uniform simple shear flow with shear rate $\dot\gamma$ is applied to the system.
Spatial inhomogeneities are neglected, and the sheared state is assumed to remain macroscopically uniform.
Under this assumption, it is convenient to describe particle motion in terms of the peculiar velocity 
\begin{equation}
    \bm{V}_i\equiv \bm{v}_i - \dot\gamma y_i \hat{\bm{e}}_x,
    \label{eq:def_V}
\end{equation}
measured relative to the local streaming velocity, where $\hat{\bm{e}}_x$ is the unit vector parallel to the $x$-direction.

In the following sections, we present a kinetic-theoretical description of the system and compare the results with simulation data in order to validate our theoretical predictions.
The simulations are performed using the Direct Simulation Monte Carlo (DSMC) method~\cite{Bird, Garcia, Poschel}. 
The reported results are obtained by averaging over 100 independent simulation ensembles to reduce statistical fluctuations.
We consider a system of $N=10^3$ particles distributed in a cubic box with a packing fraction $\varphi=0.01$. The system size is given by $L = [\pi N/(6\varphi)]^{1/3} d \simeq 37.41 d$.
To impose shear flow, we employ the SLLOD dynamics~\cite{Evans84, Evans} together with the Lees--Edwards boundary condition~\cite{Lees72}, which enables us to model an idealized uniform shear flow.

%%%%%%%%%%%%%%%%%%%%%%%%%%%%%%%%%
\section{Kinetic theory}\label{sec:kinetic_theory}
In the uniformly sheared regime, the single-particle velocity distribution function $f(\bm{V},t)$ obeys the Boltzmann equation~\cite{Chapman, Garzo, Santos04}
\begin{equation}
    \left(\frac{\partial}{\partial t}
    -\dot\gamma V_y \frac{\partial}{\partial V_x}\right)f(\bm{V}, t)=J(f,f),
    \label{eq:Boltzmann}
\end{equation}
where $J(f,f)$ denotes the nonlinear collision operator.
For hard-sphere interactions, it takes the form
\begin{align}
    J(f,f)
    &= d^2\int d\bm{V}_2 \int d\hat{\bm{k}}
    \Theta\left(\bm{V}_{12}\cdot \hat{\bm{k}}\right)
    \left(\bm{V}_{12}\cdot \hat{\bm{k}}\right)\nonumber\\
    %%%
    &\hspace{1em}\times
    \left[\frac{1}{e^2}f(\bm{V}_1^{\prime\prime},t)f(\bm{V}_2^{\prime\prime},t)
    - f(\bm{V}_1, t)f(\bm{V}_2,t)\right],
\end{align}
where $(\bm{V}_1^{\prime\prime}, \bm{V}_2^{\prime\prime})$ are the pre-collisional velocities associated with the post-collisional velocities $(\bm{V}_1, \bm{V}_2)$ determined from Eq.~\eqref{eq:pre_post_v}.

Although an exact solution of the Boltzmann equation under uniform shear is not available, it is well established that Grad's moment expansion provides a reliable approximation in dilute or moderately dense granular gases~\cite{Grad49, Garzo, Santos04, Takada18, Hayakawa19, Hayakawa17, Takada20, Kikuchi26, Kobayashi25, Kobayashi26}.
Within this framework, the velocity distribution function is approximated as
\begin{equation}
    f(\bm{V},t)
    =f_\mathrm{M}(\bm{V},t)
    \left[1+\frac{m}{2T}\left(\frac{P_{\alpha\beta}}{nT}-\delta_{\alpha\beta}\right)V_\alpha V_\beta\right],
    \label{eq:Grad}
\end{equation}
where 
\begin{equation}
    f_\mathrm{M}(\bm{V}, t)
    \equiv n\left(\frac{m}{2\pi T}\right)^{3/2}
    \exp\left(-\frac{mV^2}{2T}\right),
\end{equation}
is the local Maxwellian distribution, and the pressure (stress) tensor is defined by 
\begin{equation}
    P_{\alpha\beta}
    \equiv \int d\bm{V}mV_\alpha V_\beta f(\bm{V}, t).
\end{equation}
This approximation has been extensively validated for dilute or moderately dense granular gases under uniform shear~\cite{Garzo, Santos04, Takada18, Hayakawa19, Hayakawa17, Takada20, Kikuchi26, Kobayashi25, Kobayashi26}, where it provides a qualitatively correct description of stress anisotropy and transport properties.
While quantitative deviations may arise in strongly nonequilibrium regimes, the Grad approximation is expected to capture the essential features of the coupled evolution of temperature and stress relevant to the present analysis.

To characterize the rheological response of the granular gas under shear, we focus on the time evolution of the kinetic contribution to the stress tensor, defined as the second velocity moment of the single-particle distribution function.
In general, the total stress tensor consists of kinetic and collisional parts; however, in the dilute limit considered here, the collisional transfer contribution is subdominant and can be safely neglected.

Taking the second velocity moment of the Boltzmann equation~\eqref{eq:Boltzmann}, one obtains the evolution equation for the stress tensor~\cite{Garzo, Santos04, Takada18, Hayakawa19, Hayakawa17, Takada20, Kikuchi26, Kobayashi25, Kobayashi26}
\begin{equation}
    \frac{\partial P_{\alpha\beta}}{\partial t}
    + \dot\gamma \left(\delta_{\alpha x}P_{y\beta}
    + \delta_{\beta x}P_{y\alpha}\right)
    = -\Lambda_{\alpha\beta},
    \label{eq:P_evol}
\end{equation}
where
\begin{equation}
    \Lambda_{\alpha\beta}
    \equiv -m\int d\bm{V} V_\alpha V_\beta 
    J(f,f),
    \label{eq:def_Lambda}
\end{equation}
represents the collisional relaxation term arising from inelastic binary collisions.

For convenience, we introduce the granular temperature $T$ and the temperature anisotropy $\Delta T$, defined respectively by 
\begin{equation}
    T\equiv \frac{P_{xx}+P_{yy}+P_{zz}}{3n},\quad
    \Delta T\equiv \frac{P_{xx}-P_{yy}}{n},
\end{equation}
where the number density is given by $n=6\varphi/(\pi d^3)$.
The granular temperature measures the isotropic part of the kinetic energy, while $\Delta T$ quantifies the anisotropy induced by shear.

Using these definitions, Eq.~\eqref{eq:P_evol} can be reduced to a closed set of coupled evolution equations for $T$, $\Delta T$, and the shear stress $P_{xy}$:
\begin{subequations}\label{eq:dyn_eqs}
\begin{align}
    &\frac{\partial T}{\partial t}
    = -\frac{2}{3n}\dot\gamma P_{xy}
    - \zeta T,\label{eq:evol_T}\quad
    %%%
    \frac{\partial \Delta T}{\partial t}
    = -\frac{2}{n}\dot\gamma P_{xy}
    - \nu \Delta T,\\
    %%%
    &\frac{\partial P_{xy}}{\partial t}
    = -\dot\gamma n\left(T-\frac{1}{3}\Delta T\right)
    - \nu P_{xy}.\label{eq:evol_Pxy}
\end{align}
\end{subequations}
Here, $\zeta$  denotes the collisional energy dissipation rate, whereas $\nu$ plays the role of an effective collision frequency controlling the relaxation of stress anisotropy.
For the velocity-dependent restitution coefficient introduced in Eq.~\eqref{eq:def_COR}, explicit expressions for $\zeta$ and $\nu$ are given by~\cite{Kikuchi26}
\begin{subequations}
\begin{align}
    \zeta
    &\equiv \frac{2\sqrt{2\pi}}{3}nd^2 v_\mathrm{T} \left[\left(1-e_1^2\right)
    + \left(e_1^2-e_2^2\right)\left(1+x\right)e^{-x}\right],\\
    %%%
    \nu
    &\equiv \frac{2\sqrt{2\pi}}{5}nd^2 v_\mathrm{T}
    \Bigg\{\left(1+e_1\right)
    \left(3-e_1\right)
    + \left(e_1-e_2\right)\nonumber\\
    %%%
    &\hspace{1em}\times
    \left[\left(e_1+e_2-2\right)
    \left(1+x\right)
    + \frac{2}{3}\left(e_1+e_2\right)
    x^2\right]e^{-x}\Bigg\},
\end{align}
\end{subequations}
with $v_\mathrm{T}\equiv \sqrt{2T/m}$ is the thermal velocity and $x\equiv mv_\mathrm{c}^2/(4T)$ is a dimensionless parameter characterizing the crossover between the two restitution regimes.
The shear viscosity $\eta$ is defined through the constitutive relation 
\begin{equation}
    \eta\equiv -\frac{P_{xy}}{\dot\gamma},
    \label{eq:def_eta}
\end{equation}
which, within the present approximation, yields
\begin{equation}
    \eta=nT\frac{\nu-\zeta}{\nu^2}.
    \label{eq:eta}
\end{equation}
Accordingly, the temporal evolution of the rheological properties of the granular gas is fully determined by solving the coupled equations~\eqref{eq:dyn_eqs}.

%%%%%%%%%%%%%%%%%%%%%%%%%%%%%%%%%
\section{Mpemba protocol}\label{sec:protocol}
In the following, we investigate the temporal evolution of the granular temperature starting from two distinct initial conditions and examine whether their relaxation curves cross at a finite time, which would signal the occurrence of the Mpemba effect.

The first protocol starts from a uniformly sheared steady state.
Specifically, for $t<0$ the system is subjected to a constant shear rate $\dot\gamma_\mathrm{ini}$ and has reached a steady state.
At $t=0$, the shear rate is suddenly changed to a target value $\dot\gamma_\mathrm{tar}$, after which the system relaxes toward the corresponding steady state associated with $\dot\gamma_\mathrm{tar}$.

The second protocol begins from an unsheared, isotropic state.
For $t<0$, the system undergoes homogeneous cooling in the absence of external driving.
At the instant $t=0$, when the temperature reaches a prescribed value $T_\mathrm{ini}$, a uniform shear flow with shear rate $\dot\gamma_\mathrm{tar}$ is imposed, and the system subsequently relaxes toward the same final steady state as in the first protocol.

To distinguish these two initial conditions, we refer to the former as the \textit{From Sheared state} (FS) protocol and the latter as the \textit{From Isotropic state} (FI) protocol.
Throughout the following, quantities associated with these protocols are labeled accordingly.
For the FS protocol, the steady state realized at $t<0$ is characterized by a relation between the initial shear rate $\dot\gamma_\mathrm{ini}$ and the corresponding steady temperature $T_\mathrm{ini}^{(\mathrm{FS})}$,
\begin{equation}
    \dot\gamma_\mathrm{ini}
    =\left.\nu \sqrt{\frac{3}{2}\frac{\zeta}{\nu-\zeta}}\right|_{T=T_\mathrm{ini}^{(\mathrm{FS})}},
    \label{eq:gd_FS}
\end{equation}
which follows from the stationary solution of Eqs.~\eqref{eq:dyn_eqs}.
This equation determines the shear rate corresponding to a given steady temperature $T_\mathrm{ini}^{(\mathrm{FS})}$.
When inverted, it may admit multiple temperature solutions for a single shear rate, depending on the parameter values.
Among these branches, those satisfying $\partial T/\partial \dot\gamma<0$ are linearly unstable and therefore physically irrelevant.
A detailed discussion of this multiplicity and stability can be found in Ref.~\cite{Kikuchi26}.
The associated anisotropic temperature and shear stress in the FS steady state are given by
\begin{equation}
    \Delta T_\mathrm{ini}^{(\mathrm{FS})}
    =\left.\frac{3\zeta}{\nu} T\right|_{T=T_\mathrm{ini}^{(\mathrm{FS})}},\ 
    %%%
    P_{xy,\mathrm{ini}}^{(\mathrm{FS})}
    = -\left.\frac{nT}{\nu}\sqrt{\frac{3}{2}\zeta\left(\nu-\zeta\right)}\right|_{T=T_\mathrm{ini}^{(\mathrm{FS})}}.
    \label{eq:FS}
\end{equation}
In contrast, the FI protocol is initialized from an isotropic state with no shear-induced anisotropy.
Accordingly, the initial conditions at $t=0$ are
\begin{equation}
    T_\mathrm{ini}^{(\mathrm{FI})}>0,\quad
    \Delta T_\mathrm{ini}^{(\mathrm{FI})}
    = P_{xy,\mathrm{ini}}^{(\mathrm{FI})}=0.
    \label{eq:FI}
\end{equation}

Here, it should be noted that, in the FI protocol, the initial shear stress satisfies $P_{xy,\mathrm{ini}}^{(\mathrm{FI})}=0$, which implies that the viscosity $\eta$ is initially zero.
In general, however, a finite viscosity exists even in the homogeneous cooling state~\cite{Brilliantov}. 
This viscosity arises, for an isotropic and spatially uniform system, as a response to small spatial fluctuations such as those in the density field. 
It can be evaluated, for example, by means of the Chapman--Enskog method~\cite{Brilliantov, Chapman}.

On the other hand, the viscosity considered in the present study is defined around a uniform shear flow and is given by Eq.~\eqref{eq:def_eta}. 
These two viscosities have different physical origins, and care must be taken not to confuse them.

The final steady state reached in both protocols is determined by the target shear rate $\dot\gamma_\mathrm{tar}$.
Its temperature $T_\mathrm{tar}$, anisotropic temperature, and shear stress are obtained from Eqs.~\eqref{eq:gd_FS} and~\eqref{eq:FS} by replacing $T_\mathrm{ini}^{(\mathrm{FS})}$ with $T_\mathrm{tar}$.
Both FS and FI systems relax toward this same steady state in the long-time limit $t\to\infty$.

The two protocols and their corresponding initial and final conditions are summarized in Table~\ref{table:protocol}.
In the following analysis, we vary $\dot\gamma_\mathrm{ini}$ for the FS protocol and $T_\mathrm{ini}^{(\mathrm{FI})}$ for the FI protocol, and examine whether the relaxation curves of the granular temperature exhibit a crossover characteristic of the Mpemba effect.

%%%%%%%%%%%%%%%%%%%%%%%%%%%%%%
\begin{table}[htbp]
    \caption{Protocols defining the two initial conditions considered in this study.
    The FS protocol starts from a steady sheared state with shear rate $\dot\gamma_\mathrm{ini}$, while the FI protocol starts from an isotropic, unsheared state.
    The dynamics for $0 \le t < \infty$ is identical and plays the central role.}
    \centering
    \begin{tabular}{c|c|c}
    \hline
    & FS & FI \\
    \hline
    $t<0$ 
    & $\dot\gamma=\dot\gamma_\mathrm{ini}$ 
    & $\dot\gamma=0$ \\
    (initial conditions) & $T_\mathrm{ini}^{(\mathrm{FS})},\ \Delta T_\mathrm{ini}^{(\mathrm{FS})},\ P_{xy,\mathrm{ini}}^{(\mathrm{FS})}$
    & $T_\mathrm{ini}^{(\mathrm{FI})},\ \Delta T_\mathrm{ini}^{(\mathrm{FI})}=P_{xy,\mathrm{ini}}^{(\mathrm{FI})}=0$ \\
    \hline
    $0 \le t < \infty$ 
    & \multicolumn{2}{c}{$\dot\gamma=\dot\gamma_\mathrm{tar}$} \\
    \hline
    $t\to\infty$ 
    & \multicolumn{2}{c}{$T_\mathrm{tar},\ \Delta T_\mathrm{tar},\ P_{xy,\mathrm{tar}}$} \\
    \hline
    \end{tabular}
    \label{table:protocol}
\end{table}
%%%%%%%%%%%%%%%%%%%%%%%%%%%%%%

%%%%%%%%%%%%%%%%%%%%%%%%%%%%%%%%%
\section{Results}\label{sec:results}
\begin{figure}[htbp]
    \centering
    \includegraphics[width=\linewidth]{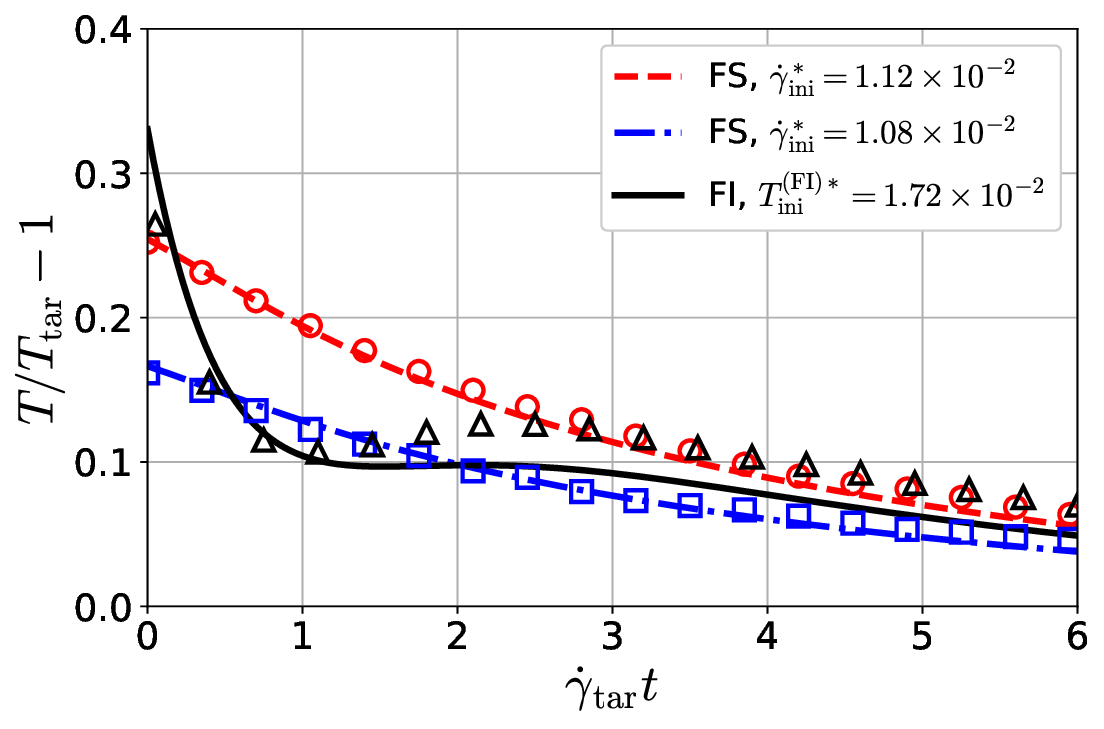}
    \caption{(Coler online) Typical time evolutions of the granular temperature for systems initialized from the FS and FI protocols.
    The FS curves correspond to $\dot\gamma_\mathrm{ini}^*=1.08\times 10^{-2}$ and $1.12\times10^{-2}$, while the FI curve corresponds to $T_\mathrm{ini}^{(\mathrm{FI})}=1.72\times 10^{-2}$, where we have introduced $\dot\gamma^*\equiv \dot\gamma d/v_\mathrm{c}$ and $T^*\equiv T/(mv_\mathrm{c}^2)$.
    The parameters are $e_1=0.50$, $e_2=0.99$, and $\dot\gamma_\mathrm{tar}^*=1.0\times 10^{-2}$.
    All quantities with an asterisk are nondimensionalized by the particle mass $m$, diameter $d$, and characteristic velocity $v_\mathrm{c}$.
    Symbols represent the DSMC results.
    }
	\label{fig:evol_T}
\end{figure}

Figure~\ref{fig:evol_T} displays typical temporal evolutions of the granular temperature for systems initialized from the FS and FI protocols.
The two curves clearly cross at a finite time, demonstrating the Mpemba effect.
For the FS protocol, the initial steady state at $t<0$ is characterized by a negative shear stress ($P_{xy}<0$), so that immediately after the shear-rate switch ($t\gtrsim 0$), the energy-injection term in Eq.~\eqref{eq:evol_T} becomes positive while collisional dissipation remains negative.
This competition suppresses the initial relaxation rate.

%%%%%%%%%%%%%%%%%%%%%%%%%%%%%%
%%%%%%%%%%%%%%%%%%%%%%%%%%%%%%

In contrast, for the FI protocol, the initial condition satisfies $P_{xy}=0$, so that the early-time dynamics is dominated by dissipation, leading to a faster initial relaxation.
As a result, when $T_\mathrm{ini}^{(\mathrm{FI})}\gtrsim T_\mathrm{ini}^{(\mathrm{FS})}$, the temperature curves cross, giving rise to the Mpemba effect.

After the initial stage, the FI protocol exhibits a nonmonotonic evolution with an overshoot before reaching the steady state.
This behavior follows from Eq.~\eqref{eq:evol_Pxy}: initially, the production term dominates and drives $P_{xy}$ in the negative direction, while at later times the relaxation term proportional to $-\nu P_{xy}$ compensates it, leading to an extremum and subsequent relaxation.
Physically, the Mpemba effect originates from the difference in the initial balance between energy injection and dissipation in the two protocols.

%%%%%%%%%%%%%%%%%%%%%%%%%%%%%%
\begin{figure}[htbp]
    \centering
    \includegraphics[width=\linewidth]{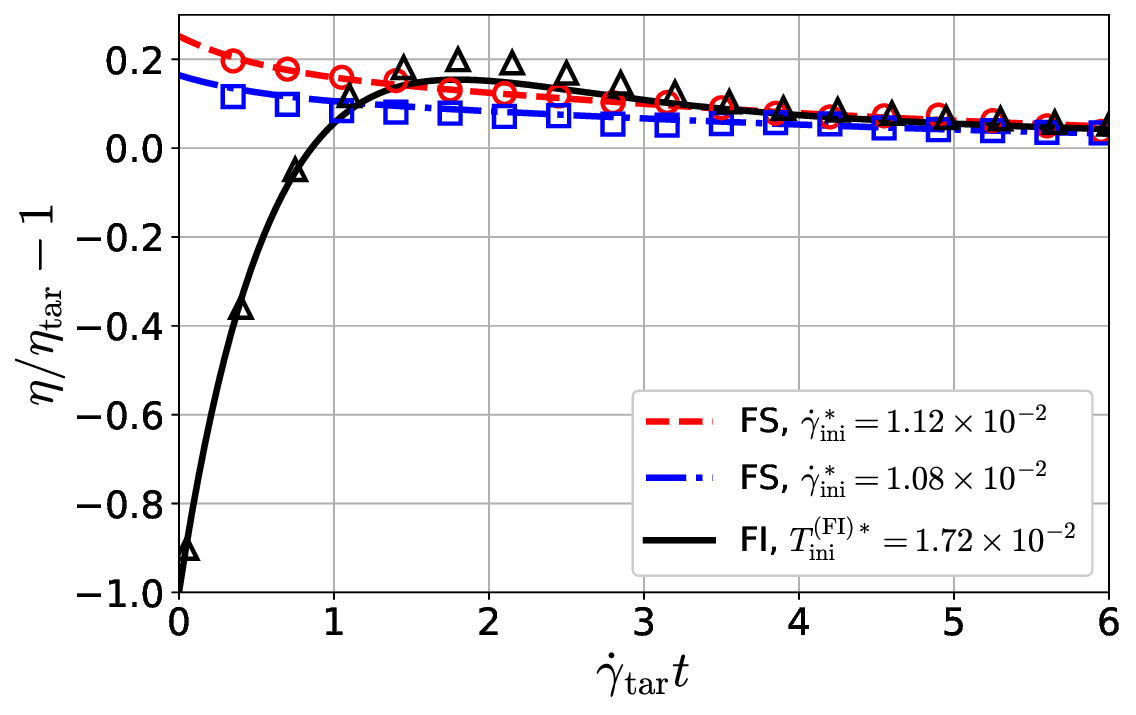}
    \caption{(Coler online) Typical time evolutions of the viscosity for the FS and FI protocols.
    The sets of parameters are identical to those used in Fig.~\ref{fig:evol_T}.
    Symbols represent the DSMC results.}
	\label{fig:evol_eta}
\end{figure}
%%%%%%%%%%%%%%%%%%%%%%%%%%%%%%

This interpretation is corroborated by the temporal evolution of the viscosity $\eta$ shown in Fig.~\ref{fig:evol_eta}.
The viscosity curves for the FS and FI protocols also cross, demonstrating a viscosity Mpemba effect~\cite{Takada21} in addition to the temperature one.

While multiple Mpemba effects have been discussed in inertial suspensions~\cite{Takada21} and quantum systems, granular gases with a constant restitution coefficient typically exhibit only a single crossover.
In the present system, the velocity-dependent restitution coefficient introduces an additional characteristic timescale associated with $v_\mathrm{c}$, which enables multiple crossings during relaxation.
We also note that the present parameter choice satisfies $e_1 < e_2$, which is associated with a non-monotonic rheological response (see Sec.~\ref{sec:discussion}).

%%%%%%%%%%%%%%%%%%%%%%%%%%%%%%
\begin{figure}[htbp]
    \centering
    \includegraphics[width=\linewidth]{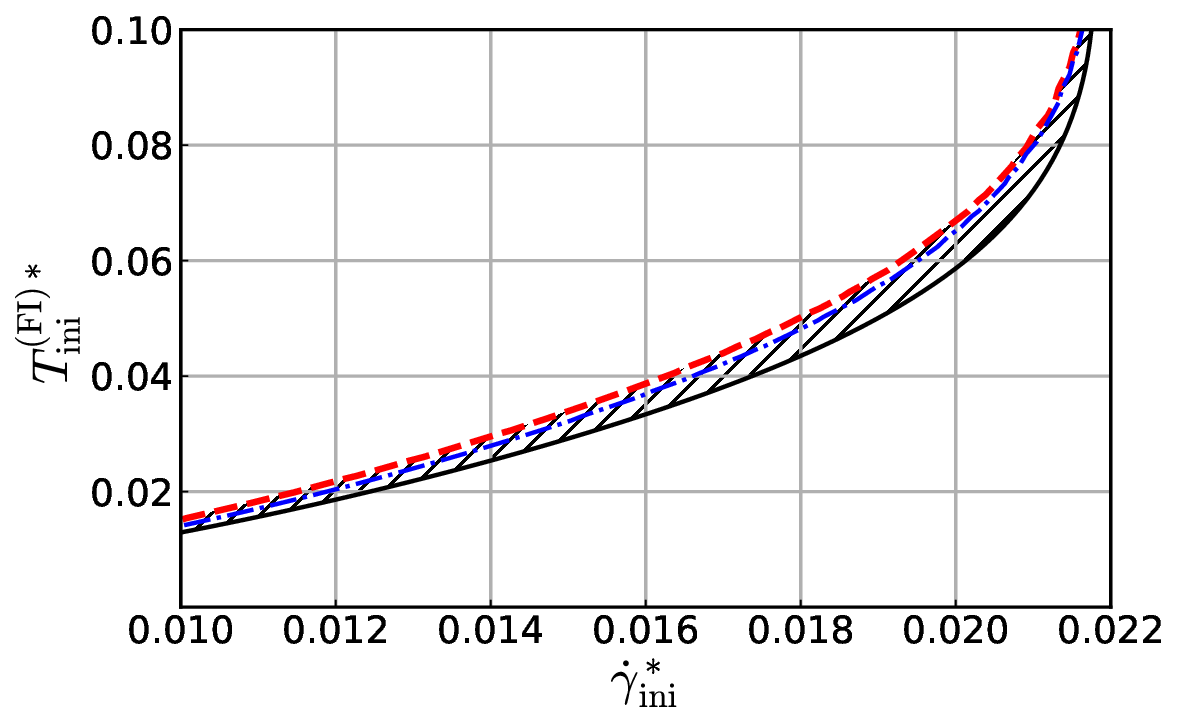}
    \caption{(Coler online) Phase diagram of the Mpemba effect for $e_1=0.50$, $e_2=0.99$, and $\dot\gamma_\mathrm{tar}^*=1.0\times 10^{-2}$.
    The shaded region indicates the parameter range where the Mpemba effect occurs.
    The region enclosed by the dashed and dotted lines corresponds to the regime in which the Mpemba effect appears twice.
    The solid line represents the theoretical prediction given by Eq.~\eqref{eq:gd_FS}.}
	\label{fig:phase_T}
\end{figure}
%%%%%%%%%%%%%%%%%%%%%%%%%%%%%%

Figure~\ref{fig:phase_T} shows the phase diagram of the temperature Mpemba effect in the parameter space spanned by the initial shear rate $\dot\gamma_\mathrm{ini}$ and the initial temperature $T_\mathrm{ini}^{(\mathrm{FI})}$.
The Mpemba effect occurs only in the region where $T_\mathrm{ini}^{(\mathrm{FI})}\gtrsim T_\mathrm{ini}^{(\mathrm{FS})}$.
When $T_\mathrm{ini}^{(\mathrm{FI})}\gg T_\mathrm{ini}^{(\mathrm{FS})}$, however, the relaxation of the FI system becomes too slow to overtake that of the FS system, and the Mpemba effect disappears.
The figure also highlights a parameter region in which the Mpemba effect occurs twice, corresponding to two distinct crossings of the temperature trajectories.
The solid line represents Eq.~\eqref{eq:gd_FS}, which separates different steady-state branches.

%%%%%%%%%%%%%%%%%%%%%%%%%%%%%%
\begin{figure}[htbp]
    \centering
    \includegraphics[width=\linewidth]{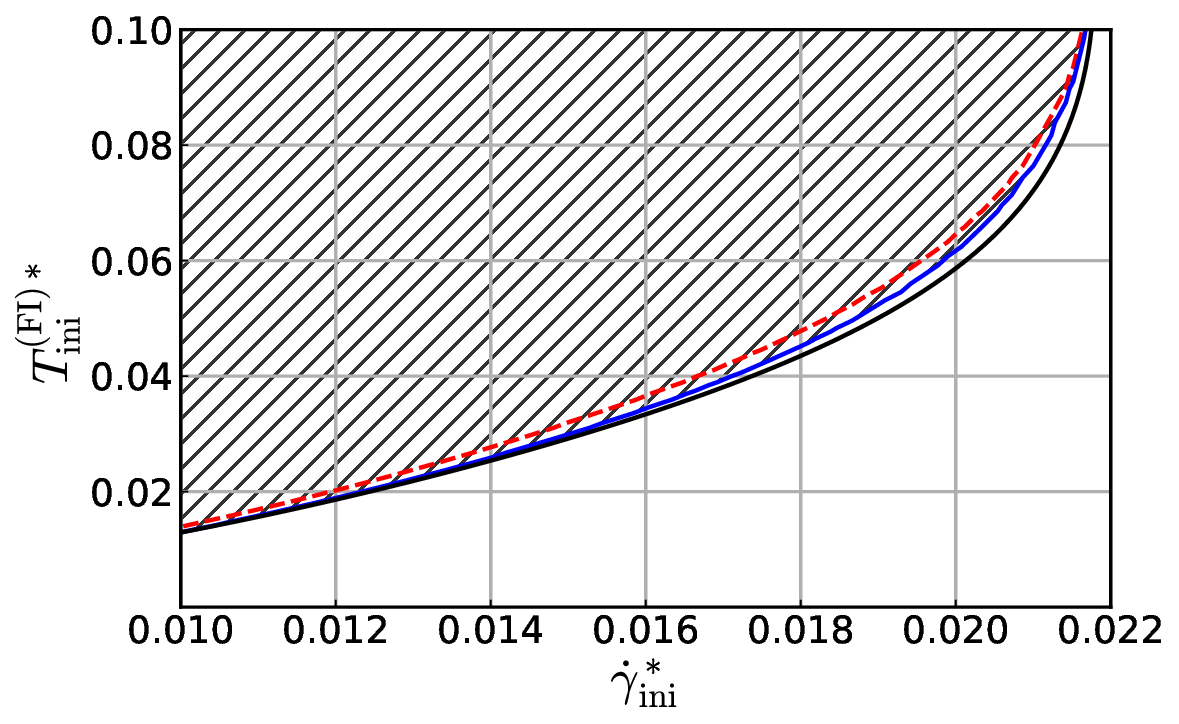}
    \caption{(Coler online) Phase diagram of the viscosity Mpemba effect.
    The shaded region indicates the parameter range where the viscosity Mpemba effect occurs.
    The region enclosed by the dashed and dotted lines corresponds to the regime in which the Mpemba effect appears twice.
    The solid line represents the theoretical prediction given by Eq.~\eqref{eq:eta} with Eq.~\eqref{eq:gd_FS}.}
	\label{fig:phase_eta}
\end{figure}
%%%%%%%%%%%%%%%%%%%%%%%%%%%%%%

Finally, Fig.~\ref{fig:phase_eta} shows the phase diagram for the viscosity Mpemba effect.
The region where the effect occurs is significantly broader than in the temperature case.
This reflects the fact that, in the FI protocol, the viscosity starts from zero and develops during relaxation, leading to a stronger transient response and facilitating crossings.
Multiple crossings are also observed in a finite parameter region.

%%%%%%%%%%%%%%%%%%%%%%%%%%%%%%%%%
\section{Discussion}\label{sec:discussion}
In the present study, the shear rate is changed discontinuously at $t=0$ from $\dot\gamma=\dot\gamma_\mathrm{ini}$ or $0$ to $\dot\gamma_\mathrm{tar}$. 
In doing so, we assume that the velocity field of the entire system instantaneously adjusts to a uniform shear flow, as given by Eq.~\eqref{eq:def_V}.

However, in realistic situations such as experiments or simulations, a finite time scale is required for the system to relax to a uniform shear flow after the boundary conditions driving the shear are changed. 
During this process, the system generally passes through spatially inhomogeneous states.

The dynamical equations~\eqref{eq:dyn_eqs} employed in this study are based on a kinetic-theoretical description under the assumption of a uniform shear flow. 
In this sense, the present protocol corresponds to an idealized limit in which spatially uniform states are realized sufficiently rapidly, or equivalently, to a reduced description where only the uniform mode is retained.
Such a treatment is expected to be valid when the relaxation of spatial fluctuations and velocity fields occurs on time scales much shorter than those of the temperature and stress evolution.

To rigorously assess the validity of this assumption, it would be necessary to investigate, in a simpler model system, such as a granular gas with a velocity-independent restitution coefficient, the relaxation of spatial inhomogeneities and the establishment of a uniform shear flow in detail. 
However, we consider this to be an independent problem in its own right and leave it for future work.

We now comment on the role of the velocity-dependent restitution coefficient in the emergence of the multiple Mpemba effect.

In the present model, the restitution coefficient changes discontinuously depending on whether the relative velocity at collision is smaller or larger than a threshold $v_{\mathrm c}$. 
This introduces an additional characteristic time scale associated with the change in dissipation mechanism, which can lead to non-monotonic relaxation dynamics and, consequently, multiple crossings of the relaxation curves.

An important observation is that the present parameter choice satisfies $e_1 < e_2$. 
From our results, as well as related study~\cite{Kikuchi26}, the multiple Mpemba effect tends to emerge in parameter regimes where the steady-state rheology exhibits a non-monotonic (S-shaped) dependence on the shear rate. 
Within the range we have examined, such non-monotonic behavior appears for $e_1 < e_2$, suggesting that this inequality plays a key role in enabling the effect.

Physically, the Mpemba effect originates from the coupling between temperature and other internal degrees of freedom, such as stress and temperature anisotropy. 
In the present system, the velocity-dependent restitution coefficient further enhances this mechanism by introducing additional time scales and non-Maxwellian features in the velocity distribution, which together allow for multiple crossings. 
A systematic investigation of the parameter dependence, including the case $e_1 > e_2$, remains an interesting open problem.

%%%%%%%%%%%%%%%%%%%%%%%%%%%%%%%%%
\section{Summary}\label{sec:summary}
The present results provide a minimal and analytically tractable example in which multiple Mpemba effects arise from purely kinetic mechanisms under shear.
We have investigated the Mpemba effect in a dilute sheared granular gas with a velocity-dependent restitution coefficient using kinetic theory. 
By analyzing the relaxation dynamics following a sudden change in the shear rate, we have demonstrated that a system starting from an isotropic state can relax faster in temperature than a system initially prepared in a sheared steady state, despite having a higher initial temperature.

We have further shown that the Mpemba effect extends to rheological properties, leading to a viscosity Mpemba effect characterized by crossings in the relaxation curves of the shear viscosity. 
Remarkably, multiple crossings can occur due to the presence of an additional intrinsic timescale introduced by the velocity dependence of the restitution coefficient. 
This mechanism is purely kinetic and does not rely on dense packing, frictional contacts, or jamming effects.

These findings establish a direct link between anomalous relaxation and nonequilibrium rheology, and suggest that controlled Mpemba phenomena can be realized in granular flows by tuning microscopic dissipation mechanisms.

%%%%%%%%%%%%%%%%%%%%%%%%%%%%%%%%%%%%%%%%%%%%%%%%%%
\begin{acknowledgment}
This work is partially supported by the Grant-in-Aid of MEXT for Scientific Research (Grant No.~JP24K06974, No.~JP24K07193, and No.~JP25K01063).
\end{acknowledgment}

%%%%%%%%%%%%%%%%%%%%%%%%%%%%%%%%%%%%%%%%%%%%%%%%%%

\end{document}